\documentstyle[12pt,epsf,epsfig]{article}
\newcount\timecount
\newcount\hours \newcount\minutes  \newcount\temp \newcount\pmhours

\hours = \time
\divide\hours by 60
\temp = \hours
\multiply\temp by 60
\minutes = \time
\advance\minutes by -\temp
\def\hour{\the\hours}
\def\minute{\ifnum\minutes<10 0\the\minutes
            \else\the\minutes\fi}
\def\clock{
\ifnum\hours=0 12:\minute\ AM
\else\ifnum\hours<12 \hour:\minute\ AM
      \else\ifnum\hours=12 12:\minute\ PM
            \else\ifnum\hours>12
                 \pmhours=\hours
                 \advance\pmhours by -12
                 \the\pmhours:\minute\ PM
                 \fi
            \fi
      \fi
\fi
}

\def\monthname{\relax\ifcase\month 0/\or January\or February\or
   March\or April\or May\or June\or July\or August\or September\or
   October\or November\or December\else\number\month/\fi}

\def\bold#1{\setbox0=\hbox{$#1$}%
     \kern-.025em\copy0\kern-\wd0
     \kern.05em\copy0\kern-\wd0
     \kern-.025em\raise.0433em\box0 }

\def\ga{\mathrel{\raise.3ex\hbox{$>$\kern-.75em\lower1ex\hbox{$\sim$}}}}
\def\la{\mathrel{\raise.3ex\hbox{$<$\kern-.75em\lower1ex\hbox{$\sim$}}}}
\def\gev{{\rm \, Ge\kern-0.125em V}}
\def\tev{{\rm \, Te\kern-0.125em V}}
\def\beq{\begin{equation}}
\def\eeq{\end{equation}}

\def\m12{m_{1\!/2}}

\def\iso#1#2{\mbox{${}^{#2}{\rm #1}$}}
\def\he#1{\iso{He}{#1}}
\def\li#1{\iso{Li}{#1}}
\def\ga{\mathrel{\raise.3ex\hbox{$>$\kern-.75em\lower1ex\hbox{$\sim$}}}}
\def\la{\mathrel{\raise.3ex\hbox{$<$\kern-.75em\lower1ex\hbox{$\sim$}}}}
\def\gyr{{\rm \, G\kern-0.125em yr}}
\def\gev{{\rm \, Ge\kern-0.125em V}}
\def\tev{{\rm \, Te\kern-0.125em V}}
\def\beq{\begin{equation}}
\def\eeq{\end{equation}}

\def\m12{m_{1\!/2}}

\newcommand {\etal}{{et\thinspace al.} }

\def\gappeq{\mathrel{\rlap {\raise.5ex\hbox{$>$}}
{\lower.5ex\hbox{$\sim$}}}}

\def\lappeq{\mathrel{\rlap{\raise.5ex\hbox{$<$}}
{\lower.5ex\hbox{$\sim$}}}}

\def\Toprel#1\over#2{\mathrel{\mathop{#2}\limits^{#1}}}

\begin{document}
\begin{titlepage}
\pagestyle{empty}
\baselineskip=21pt
\rightline{astro-ph/0503023}
\rightline{CERN-PH-TH/2005-030}
\rightline{UMN--TH--2345/05}
\rightline{FTPI--MINN--05/04}
\vskip 1in
\begin{center}
{\large{\bf The Effects of Unstable Particles on Light-Element Abundances: 
Lithium versus Deuterium and $^3$He 
 %Reconciling the BBN/WMAP-Predicted $^7$Li Abundance with Observations
 }}
\end{center}
\begin{center}
\vskip 0.2in
{{\bf John Ellis}$^1$, {\bf Keith
A.~Olive}$^{2}$ and {\bf Elisabeth Vangioni}$^{3}$}\\
\vskip 0.1in
{\it
$^1${TH Division, Physics Department, CERN, CH-1211 Geneva 23, 
Switzerland}\\
$^2${Theoretical Physics Institute,
University of Minnesota, Minneapolis, MN 55455, USA}\\
$^3${Institut d'Astrophysique de Paris, F-75014 Paris, France}}\\
\vskip 0.2in
{\bf Abstract}
\end{center}
\baselineskip=18pt \noindent

%%%%%%%%%%%%%%%%%%%%%%%%%%%%%%%%%%%%%%%%%%%%%%%%%%%%%%%%%%%%%%%%%%%%%

We reconsider the effects of unstable particles on the production and
destruction of the primordial light elements, with a view to reconciling
the high primordial $^7$Li abundance deduced from Big Bang Nucleosynthesis
(BBN), as implied by the baryon-to-photon ratio now inferred from the
anisotropies of the Cosmic Microwave Background (CMB), with the lower
abundance of $^7$Li observed in halo stars. The potential destruction of
$^7$Li is strongly constrained by observations of Deuterium (D), $^3$He
and $^6$Li.  We identify ranges for the unstable particle abundance and
lifetime which would deplete $^7$Li while remaining consistent with the
abundance of $^6$Li. However, in these regions either the D abundance is
unacceptably low or the ratio \he3/D is unacceptably large. We conclude
that late particle decay is unable to explain both the discrepancy of the
calculated \li7 abundance and the observed \li7 plateau. In the context of
supersymmetric theories with neutralino or gravitino dark matter, we
display the corresponding light-element constraints on the model
parameters.

%%%%%%%%%%%%%%%%%%%%%%%%%%%%%%%%%%%%%%%%%%%%%%%%%%%%%%%%%%%%%%%%%%%%%
\vfill
\leftline{CERN-PH-TH/2005-030}
%%%%%%%%%%%%%%%%%%%%%%%%%%%%%%%%%%%%%%%%%%%%%%%%%%%%%%%%%%%%%%%%%%%%%%%%%%%%%%
\end{titlepage}

\section{Introduction}

The observed abundances of light elements are generally in good agreement
with the predictions of Big Bang Nucleosynthesis (BBN) calculated assuming
a homogeneous Robertson-Walker-Friedman cosmology \cite{bbn}. Within this
framework, the success of BBN calculations imposes important constraints
on the number of light particle species and on the baryon-to-photon ratio
$\eta$ \cite{ssg}. Recent high-precision measurements of the cosmic
microwave background (CMB) radiation by WMAP and other experiments now
complement the BBN in important ways \cite{wmap}. For example, they impose
strong constraints on $\eta$ and weaker constraints on the number of light
particle species. Of particular interest are the very precise predictions
of the light element abundances from BBN that are made with the CMB value
of $\eta$~\cite{cfo3}.

Based on these predictions, there is now tension between some observed
light-element abundances and those that would be calculated using the CMB
value of $\eta$ and assuming no additional light particles beyond three
light neutrino species. In particular, the prediction for the primordial
abundance of $^7$Li made using the CMB value of $\eta$
\cite{cfo,coc,coc2,cuoco,cyburt} is somewhat higher (by a factor of 2 - 3)
than the primordial abundance inferred from astrophysical observations.
The significance of this discrepancy should not be over-emphasized, in
view of the potential systematic errors in the interpretation of the
astrophysical data. However, it has stimulated theoretical explorations of
mechanisms for modifying the CMB/BBN prediction for $^7$Li, for example
via the late decays of massive particles \cite{jed,feng}.

The effects of such decays have been studied extensively, and the
constraints imposed on them by the observed abundances of $^4$He,
Deuterium and $^6$Li are well understood \cite{decays,cefo,kawa}.  
Previous studies had shown no incompatibilities between these constraints
and the suggestion that late-decaying particles might have modified the
BBN prediction for the abundance of $^7$Li so as to agree better with the
astrophysical observations. Supersymmetric models with conserved $R$
parity naturally predict such a particle, either the gravitino if it is
not the lightest supersymmetric particle (LSP), or a neutralino or stau
slepton if the gravitino is the LSP~\cite{morefeng,eoss5}, which may well
have the appropriate abundance and lifetime to affect both the \li6 and
\li7 abundances. Later in this paper, we identify these regions in the
parameter space of a constrained version of the minimal supersymetric
extension of the Standard Model (CMSSM), as well as in very constrained
versions of the model \cite{vcmssm} motivated by supergravity
considerations, which predict a gravitino LSP in parts of the parameter
space.

However, we also point out in this paper that such scenarios typically
yield an abundance of $^3$He that may be more than an order of magnitude
larger than the Deuterium abundance. Since it appears that the
abundance of $^3$He has remained relatively constant in time, 
whereas Deuterium (D) would have been destroyed in stars, it 
seems unlikely that the ratio of the $^3$He to D abundances 
could have been significantly larger than it present value in the 
early history of the Universe~\cite{kawa}. Imposing this
constraint, we find a further restriction on the parameter space of
late-decaying heavy-particle models, which excludes the region of
parameter space where they could have the desired impact on the primordial
abundance of $^7$Li. We display this and other constraints on the lifetime
and abundance of any massive unstable relic particle such as a gravitino,
pointing out also the potential impact of weakening the lower limit on the
primordial Deuterium abundance. We also display the effects of the $^3$He
constraint on the parameter spaces of models in which heavier
supersymmetric sparticles decay into gravitinos. The overall reductions in
the allowed parameter spaces are often small, but they do exclude the
regions where the abundance of $^7$Li could be brought into line with the
astrophysical observations. This may motivate a reassessment of their
interpretation.

\section{Is there a problem with Lithium-7?}

The most direct and accurate estimate of the baryon-to-photon ratio 
$\eta$ is currently provided by the acoustic structures in the CMB 
perturbations \cite{wmap}, namely
\begin{equation}
\eta \; = \; 6.14 \pm 0.25 \times 10^{-10}.
\label{BBNeta}
\end{equation}
This range may be used an input into homogeneous BBN calculations 
\cite{cfo,coc,coc2,cuoco,cyburt}, 
yielding the following abundances for the elements of 
principal interest, which are taken from~\cite{cyburt}:
\begin{eqnarray}
Y_p  \; & = & \; 0.2485 \pm 0.0005, \nonumber \\
{{\rm D} \over {\rm H}} \; & = & \; 2.55^{+0.21}_{-0.20} \times 10^{-5},
\nonumber \\
{^3{\rm He} \over {\rm H}} \; & = & \; 1.01\pm 0.07 \times 10^{-5},
\nonumber \\
{^7{\rm Li} \over {\rm H}} \; & = & \; 4.26^{+0.73}_{-0.60} \times 
10^{-10}, \nonumber \\
{^6{\rm Li} \over {\rm H}} \; & = & \; 1.3 \pm 0.1 \times 10^{-14},
\label{BBNcalx}
\end{eqnarray}
where $Y_p$ is the \he4 mass fraction, and the other abundances are 
expressed in terms of their numbers relative to H, as shown.

These abundances may then be compared with the abundances of the same
elements that are inferred from those observed in the most primitive
astrophysical sites~\cite{OlSk,kirkman,rbofn,Asplund2,Lambert}:
\begin{eqnarray}
Y_p \; & = & \; 0.232~~ {\rm to}~~  0.258, \nonumber \\
{{\rm D} \over {\rm H}} \; & = & \; 2.78 \pm 0.29 \times 10^{-5}, 
\nonumber \\
{^3{\rm He} \over {\rm H}} \; & = & \; 1.5 \pm 0.5 \times 
10^{-5}, \nonumber\\
{\li7 \over {\rm H}} \; & = & \; 1.23^{+0.68}_{-0.32} \times 10^{-10},
\nonumber\\
{^6{\rm Li} \over {\rm H}} \; & = & \; 6^{+7}_{-3} \times 
10^{-12}.
\label{observed}
\end{eqnarray}
Note that, for $^3$He, we have at our disposal only local data coming from
star-forming HII regions in the galactic disk~\cite{Bania} or from the proto-solar
value \cite{gg}. Comparing the
two sets of abundances, (\ref{BBNcalx}) and (\ref{observed}), we see no
significant discrepancies, except in the case of $^7$Li and $^6$Li.  
However, whereas $^7$Li has to be destroyed, one must produce a factor 
1000 more $^6$Li.
 
The value quoted above for the \li7 abundance assumes that Li depletion is
negligible in the stars observed. Indeed, standard stellar evolution
models predict Li depletion factors which are very small (less than 0.05
dex) in very metal-poor turnoff stars~\cite{ddk}. However, there is no
reason to believe that such simple models incorporate all effects which
could lead to depletion, such as rotationally-induced mixing and/or 
diffusion.
Including these effects, current estimates for possible depletion factors
are in the range $\sim$~0.2--0.4~dex~\cite{dep}.  However, the data
sample~\cite{rnb} used in deriving the abundance in (\ref{observed}) shows
a negligible intrinsic spread in Li, leading to the conclusion that
depletion in these stars is in fact quite low (less than 0.1 dex).
 
Another important source of potential systematic uncertainty is related
to the assumed surface temperature of the star.  A recent study~\cite{mr}
found significantly and systematically higher temperatures for
stars used in \li7 observations, specifically at low metallicity. 
This result leads to \li7/H = $(2.34 \pm
0.32) \times 10^{-10}$, which is still, however, nearly a factor of 2
smaller than the BBN/WMAP prediction. We note, finally, that another
potential source for theoretical uncertainty lies in the BBN calculation 
of
the \li7 abundance.  However, this too has been shown to be incapable of
resolving the \li7 discrepancy~\cite{coc,cfo4}.

The comparison of (\ref{BBNcalx}) and (\ref{observed}) also shows a
discrepancy for \li6.  It has generally been assumed that \li6 is produced
in post-BBN processes such as galactic cosmic-ray nucleosynthesis
\cite{li6}. Until recently, the abundance of $^6$Li had been observed only
in a few metal-poor halo stars with metallicity [Fe/H] larger than -2.3.
However, new observations of this isotope have now been obtained in halo
stars.  New values of the ratio $^6$Li/$^7$Li have been measured with UVES
at the VLT-UT2 Kueyen ESO telescope, in halo stars with metallicity
ranging from -2.7 to -0.5~\cite{Asplund2,Lambert}. These observations
indicate the presence of a plateau in \li6/H $\simeq\, 10^{-11}$, suggesting
a pregalactic origin for the formation of \li6~\cite{Rollinde}.  These
data provide interesting new constraints in the present context.

The relation of the observed $^7$Li abundance to its primordial value may
be debated, but for the moment we take the discrepancy between the
observed and calculated $^7$Li abundances at face value, and explore its
possible theoretical interpretation in terms of late-decaying massive
particles, depleting $^7$Li and possibly producing $^6$Li, without
negative effects on D and $^3$He.

\section{Possible Impact of Late-Decaying Particles}

The decays of massive particles $X$ with lifetimes $> 10^2$~s could, in
principle, have modified the BBN predictions in either of two ways. 
First, their
decay products would have increased the entropy in the primordial plasma,
implying that its value during BBN was lower than that inferred from the
CMB. However, this effect is negligible compared with the second effect,
which is the modification of the BBN light-element abundances by the
interactions of decay products \cite{decays}.

The latter possibility has been explored recently as a mechanism for
reducing the primordial $^7$Li abundance \cite{jed,feng}. The existence of
such late-decaying massive particles $X$ is a generic possibility in
supersymmetric models, in particular \cite{morefeng,eoss5}. Examples
include a massive gravitino weighing $\sim 100$~GeV, if it is {\it not} 
the
lightest supersymmetric particle (LSP), or some other next-to-lightest
supersymmetric particle (NSP) if the gravitino {\it is} the LSP. 
Cosmological
constraints on such scenarios have been explored 
previously~\cite{decays,cefo,kawa}. However, the
potential $^7$Li problem motivates a re-examination of the astrophysical
and cosmological constraints on such scenarios. In particular, we wish to
determine whether or not a possible solution to this problem can be found
in the context of motivated and well-studied supersymmetric models.

\subsection{The D and $^4$He Abundances}
\begin{figure}
\vskip 0.5in
\vspace*{-0.75in}
%\hspace*{-.70in}
\begin{center}
\epsfig{file=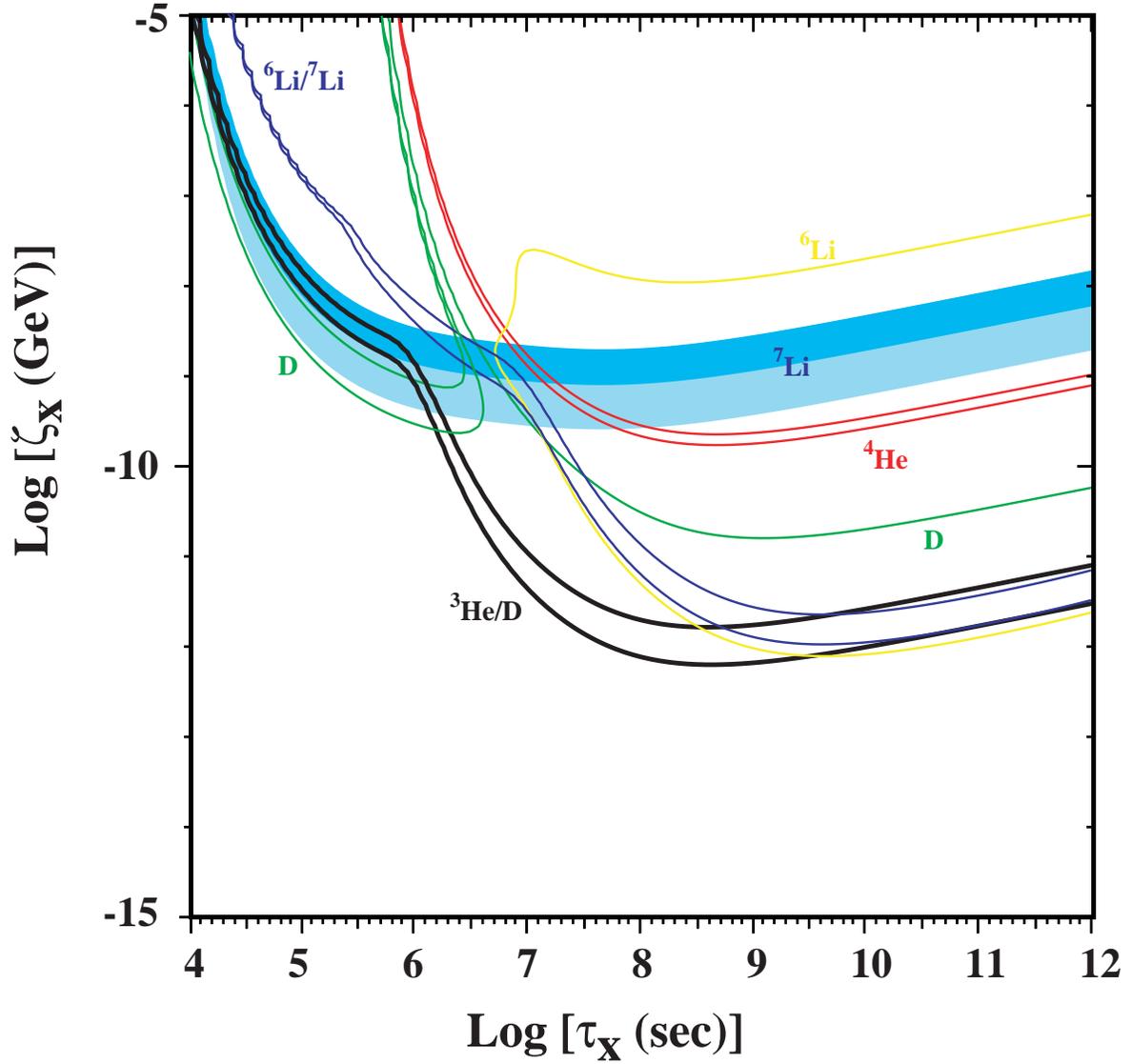,height=6in}
\hfill
\end{center}
\caption{
{\it 
The constraints imposed by the astophysical observations of $^4$He (red 
lines), $D/H$ (green lines), $^6$Li (yellow line), $^6$Li/$^7$Li (blue 
lines), $^7$Li (blue band) and $^3$He (black lines).}}
\label{fig:tauzetacleft}
\end{figure}

Fig.~\ref{fig:tauzetacleft} shows the current constraints in 
the $(\tau_X, \zeta_X)$ plane, updating those shown in~\cite{cefo}. The 
green 
lines are the contours
\begin{equation}
(1.3 \; {\rm or} \; 2.2) \times 10^{-5} \; < \; {{\rm D} \over {\rm H}} \; 
< \; 5.3 
\times 10^{-5}.
\label{Dlines}
\end{equation}
The first of the lower bounds is the higher line to the left of the cleft, 
and represents the very conservative lower limit on 
D/H assumed in~\cite{cefo}. The range 1.3 -- 5.3 $\times 10^{-5}$
effectively brackets all recent observations of D/H in quasar absorption 
systems. The second of the lower bounds is the lower line on the left 
side, and represents what now seems a 
reasonable lower bound, which is obtained from the 
2-$\sigma$ lower limit in (\ref{observed}). The upper bound in 
(\ref{Dlines}) is the line to the right of 
the cleft, and is the same upper limit as was 
used in~\cite{cefo}. 
{\it A priori}, there is also a narrow strip 
at larger $\zeta_X$ and $\tau_X$ where the D/H ratio also falls within
the range (\ref{Dlines}), but this is excluded by the 
observed $^4$He abundance.

The solid red lines in the upper right part of Fig.~\ref{fig:tauzetacleft} 
correspond to the limits
\begin{equation}
Y_p \; > \; 0.227\;{\rm or}\; 0.232,
\label{4lines}
\end{equation}
where the lower number (corresponding to the higher line) was used 
in~\cite{cefo}, and the higher number (corresponding to the lower line) 
is a lower limit that has been advocated recently \cite{OlSk}.
It is apparent that, for our purposes, the third significant figure 
in the $^4$He abundance is unimportant: the narrow D/H strip is 
in any case excluded, and there are always stronger bounds on $\zeta_X$ at 
large $\tau_X$. 

\subsection{The $^6$Li Abundance}

As said above, recent observations of $^6$Li in halo stars have provided
new insight into the origin and the evolution of this isotope. We recall
that $^6$Li is a pure product of spallation, and many studies have
followed the evolution of $^6$Li in our Galaxy~\cite{li6}. Of particular
importance in this context is the $\alpha+\alpha$ reaction that leads to
the synthesis of this isotope as well as $^7$Li, and is efficient very
early in the evolutionary history of the Galaxy.  The new values of
$^6$Li/$^7$Li that have been measured in halo stars with UVES at the
VLT-UT2 Kueyen ESO telescope indicate the presence of a plateau in $^6$Li,
which suggests a pregalactic origin for the formation of this isotope.
The evolution of $^6$Li with
redshift was calculated \cite{Rollinde} 
following an initial burst of cosmological cosmic rays up to the
formation of the Galaxy. This process is able to produce the required
abundance of $^6$Li observed in metal-poor halo stars without the
additional over-production of $^7$Li. In this paper, we have to consider
the new constraint brought by the existence of this plateau. The potential
destruction of $^7$Li by unstable particles must not lead to
over-production of $^6$Li.

The constraint imposed by the $^6$Li abundance is shown as a solid 
yellow line in Fig.~\ref{fig:tauzetacleft}, which is the same as that discussed 
in~\cite{cefo}. Also shown, as solid blue lines, are two contours 
representing possible upper limits on the $^6$Li/$^7$Li ratio:
\begin{equation}
{^6{\rm Li} \over ^7{\rm Li}} \; < \; 0.07 \; {\rm or} \; 0.15,
\label{67lines}
\end{equation}
with the upper (lower) contour corresponding to the upper (lower) number
in (\ref{67lines}). The lower number was used in \cite{cefo} and
represented the upper limit available at the time, which was essentially
based on multiple observations of a single star. The most recent data
\cite{Asplund2,Lambert} includes observations of several stars. The Li
isotope ratio for most metal-poor stars in the sample is as high as 0.15,
and we display that upper limit here.  The main new effect of this
constraint is to disallow a region in the near-vertical cleft between the
upper and lower limits on D/H, as seen in Fig.~\ref{fig:tauzetacleft}.

\subsection{The $^7$Li Abundance}

The main region of interest in Fig.~\ref{fig:tauzetacleft} is the blue
shaded band that represents the inferred $^7$Li abundance:
\begin{equation}
0.9 \times 10^{-10} \; < \; {^7{\rm Li} \over {\rm H}} \; < \; (2 \; {\rm 
or} \; 3) \times 10^{-10},
\label{7bands}
\end{equation}
with the $^7$Li abundance decreasing as $\zeta_X$ increases and the 
intensity of the shading changing at the intermediate value. In \cite{cefo},
only the lower bound was used due the existing discrepancy between
the primordial and observationally determined values. It is 
apparent that 
$^7$Li abundances in the lower part of the range (\ref{7bands}) are 
possible only high in the Deuterium cleft, and even then only if one 
uses the recent and more relaxed limit on the $^6$Li/$^7$Li ratio (\ref{67lines}). 
Values of the $^7$Li abundance in the upper part of the range 
(\ref{7bands}) are possible, however, even if one uses the more stringent 
constraint on $^6$Li/$^7$Li. In this case, the allowed region of parameter 
space would also extend to lower $\tau_X$, if one could tolerate values of 
D/H between 1.3 and $2.2 \times 10^{-5}$.

For the convenience of the subsequent discussion, the region of the 
$(\tau_X, \zeta_X)$ plane that is of interest for lowering the $^7$Li 
abundance is shown alone in panel (a) of Fig.~\ref{fig:ridiculous}. The 
blue region in the arc at low $\tau_X$ is the region excluded by the 
stronger lower limit on the Deuterium abundance: D/H $> 2.2 \times 
10^{-5}$, and the red region to its right is the extra domain that is 
excluded by the $^3$He/D ratio, as we discuss below.

\section{The Importance of the $^3$He Abundance}

\begin{figure}
\vskip 0.5in
\vspace*{-0.75in}
%\hspace*{-.70in}
\begin{minipage}{8in}
\epsfig{file=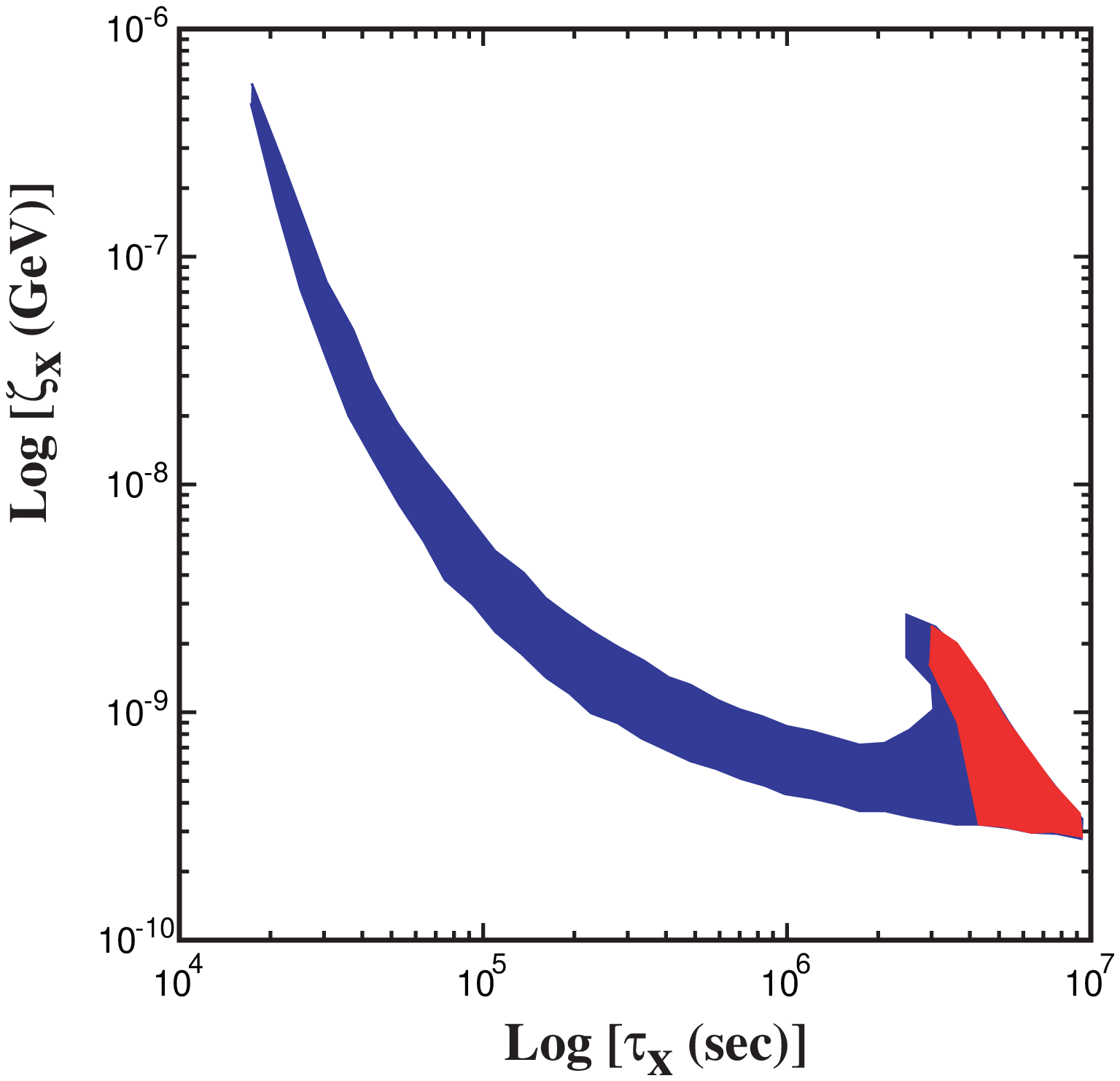,height=3.3in}
\hspace*{-0.17in}
\epsfig{file=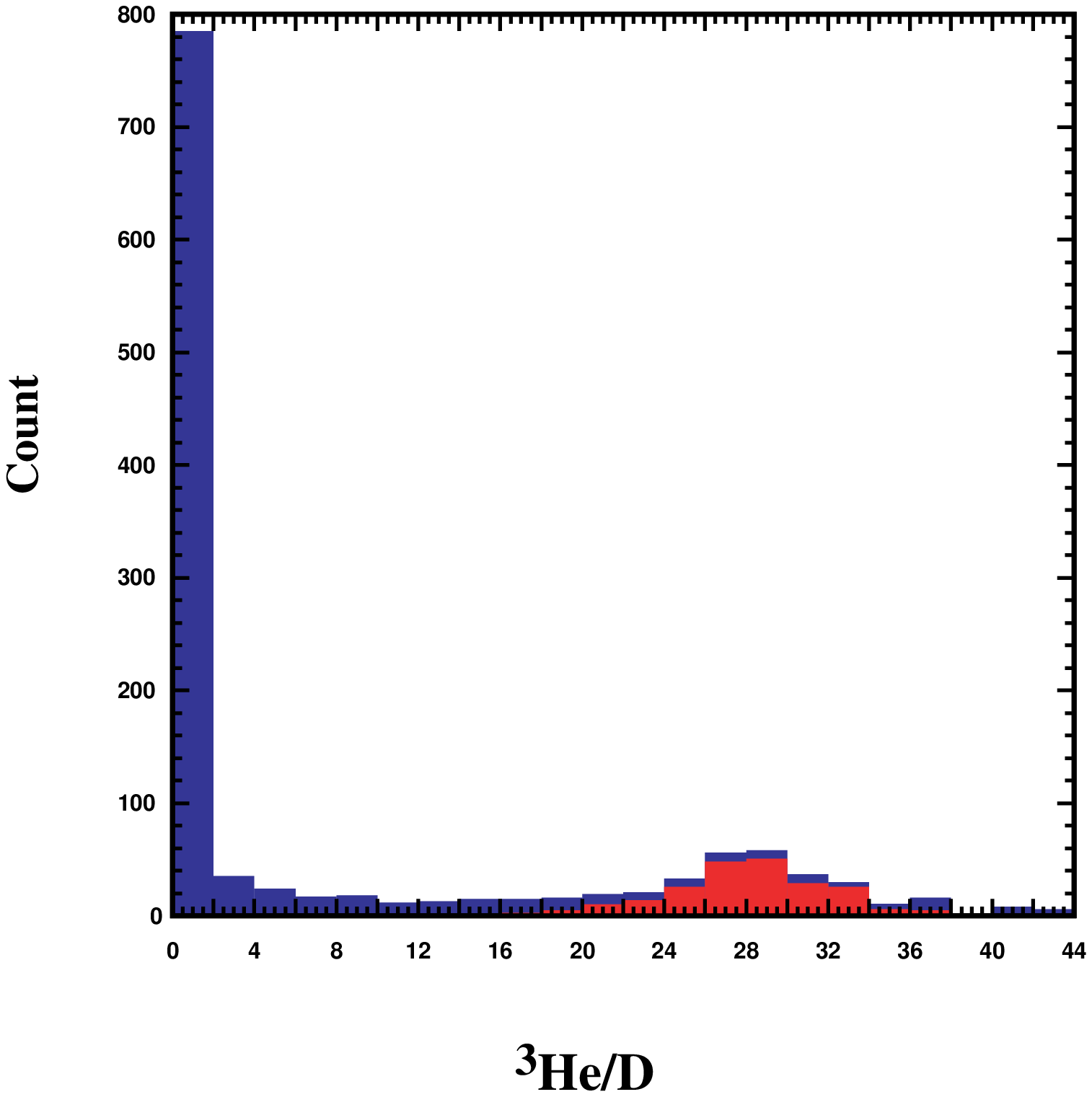,height=3.3in} 
\hfill
\end{minipage}
\caption{
{\it (a) The region of the $(\tau_X, \zeta_X)$ plane in which a decaying 
relic 
particle could have the desired impact on the $^7$Li abundance. 
To derive the blue (darker grey) region, the only abundance cuts
applied are: 
$0.9 < \li7/{\rm H} \times 10^{10} < 3.0$,
$1.3 < {\rm D}/{\rm H} \times 10^5 < 5.3$, and \li6/\li7 $< 0.15$.  To 
obtain the red (lighter grey) region, the lower
bound on ${\rm D}/{\rm H}$ was increased to $2.2 \times 10^{-5}$.  (b) A 
histogram of the $\he3/{\rm D}$ ratios found in scenarios sampling the 
region displayed in panel (a), with similar colour coding. }} 
\label{fig:ridiculous} 
\end{figure}

We now come to the constraint from \he3, namely that the $^3$He/D ratio 
is absurdly high in the Deuterium cleft. Panel (b) of 
Fig.~\ref{fig:ridiculous} shows a histogram of the values of the $^3$He/D 
ratio found in a dense sample of scenarios in the interesting regions
shown in panel (a). Since Deuterium is more fragile than $^3$He, whose 
abundance is thought to have remained roughly constant since primordial 
nucleosynthesis when comparing the BBN value to it proto-solar abundance,
one would expect, in principle,  the $^3$He/D ratio to have been 
increased by stellar processing. Indeed,  there is considerable uncertainty in the evolution
of \he3~\cite{vofc}. This uncertainty is largely associated with the degree to which \he3 is  
produced or destroyed in stars.
Since D is totally destroyed in stars, the ratio of \he3/D can only increase in time or remain constant
if \he3 is also completely destroyed in stars. 
The present or proto-solar value of 
$^3$He/D can therefore be used to set an upper limit on the primordial 
value. Fig.~\ref{fig:tauzetacleft} displays the upper limits 
\begin{equation}
{^3{\rm He} \over D} \; < \; 1 \; {\rm or} \; 2
\label{3lines}
\end{equation}
as solid black lines. Above these contours, the value of $^3$He/D
increases very rapidly, and points high in the Deuterium cleft of 
Fig.~\ref{fig:tauzetacleft} have absurdly high values of $^3$He/D, 
exceeding the limit (\ref{3lines}) by an order of magnitude or more. 
These are the red points producing the high-end peak of the histogram shown in 
panel (b) of Fig.~\ref{fig:ridiculous}, whereas the blue points are those 
excluded by the lower limit D/H $> 2.2 \times 10^{-5}$ that is now 
preferred. We see that these points mostly have acceptably low values of 
$^3$He/D, though some large values are found near the boundary with the 
red region in panel (a) of Fig.~\ref{fig:ridiculous}.

It is interesting to note that in the red region which has acceptable D/H
and a \li7/H abundance low enough to match the observed values, the \li6
abundance is relatively high: $7.3 \times 10^{-12} < $\li6 /H$ < 1.6
\times 10^{-11}$. This matches the new \li6 observations quite well, and
would circumvent the need for an early period of \li6 production by
cosmological cosmic rays. Unfortunately, \he3/D ranges from 17 -- 37 for
these parameter values.

The previous upper limit on $\eta_X$~\cite{cefo} corresponded to the
constraint $m_X n_X / n_\gamma < 5.0 \times 10^{-12}$~GeV for $\tau_X =
10^8$~s. The weaker (stronger) version of the $^3$He constraint adopted 
here corresponds to
\begin{equation}
m_X {n_X \over n_\gamma} \; < \; 2.0 (0.8) \times 10^{-12}~{\rm GeV}
\label{new8}
\end{equation}
for $\tau_X = 10^8$~s. However, the impact of the $^3$He constraint is
even stronger for $\tau_X = 10^7$~s, the location of the previous
Deuterium cleft. The analysis of~\cite{cefo} would have given $m_X n_X /
n_\gamma < 360 \times 10^{-12}$~GeV, whereas the weaker (stronger) $^3$He
constraint adopted here corresponds to
\begin{equation}
m_X {n_X \over n_\gamma} \; < \; 9.3 (3.8) \times 10^{-12}~{\rm GeV}
\label{new7}  
\end{equation}
for $\tau_X = 10^7$~s.

\section{Applications to Supersymmetric Scenarios}

We now discuss some examples of the consequences of the $^3$He constraint
for various supersymmetric scenarios in which $R$ parity is conserved. In 
such models, if the gravitino is not the LSP it will generically decay 
gravitationally with a long lifetime. If the gravitino is the LSP, the 
next-to-lightest supersymmetric particle (NSP) will decay gravitationally 
into the gravitino LSP, again with a long lifetime.

\subsection{Models with an Unstable Gravitino}

We first consider the possibility that the gravitino is not the lightest
supersymmetric particle (LSP), which is instead the lightest neutralino
$\chi$. In this case, the gravitino is unstable, with a lifetime that
could well fall within the range considered here. In such a scenario, the
light-element abundances impose an important upper limit on the possible
temperature of the Universe, e.g., during reheating after inflation, which
we denote by $T_R$~\cite{decays,cefo,kawa}. We recall that thermal reactions are 
estimated to produce an abundance of gravitinos given by~\cite{cefo}:
\begin{equation}
{n_{m_{3/2}} \over n_\gamma} \; = \; (0.7 - 2.7) \times 10^{-11} \times 
\left( {T_R \over 10^{10}~{\rm GeV}} \right).
\label{Y32}
\end{equation}
Assuming that $m_{3/2} = 100$~GeV and $\tau_X = 10^8$~s, and imposing the 
constraints (\ref{new8}), we now find
\begin{equation}
T_R \; < \; (0.8 - 2.8) \times 10^7~{\rm GeV}, \; ((0.3 - 1.1) \times 
10^7~{\rm GeV})
\label{TR}
\end{equation}
for the weaker (stronger) version of the $^3$He constraint. This becomes
an even more significant constraint on inflationary models, which were
already somewhat embarrassed by the previous upper limit $T_R \sim 2
\times 10^7$~GeV.

\subsection{Models with Gravitino Dark Matter}

We now consider the possibility that the gravitino is itself the LSP, in
which case the next-to-lightest supersymmetric particle (NSP) would
instead be unstable, also with a long lifetime. We will work in the context
of the constrained MSSM (CMSSM)~\cite{cmssmnew,eoss,cmssmmap}, so that all 
the scenarios we
consider have universal soft supersymmetry-breaking scalar masses $m_0$
and gaugino masses $m_{1/2}$ at the GUT scale before renormalization. The
magnitude of the Higgsino mixing parameter $|\mu|$ and the pseudoscalar
Higgs mass $m_A$ are fixed by the electroweak vacuum conditions. We
consider scenarios with $\mu > 0$, which are favoured by $g_\mu - 2$ and,
to a lesser extent, $b \to s \gamma$. The scenarios that we study differ
in their assumptions about the relationship of the gravitino mass
$m_{3/2}$ to $m_0$, but they all share the common feature that the LSP is
the gravitino in generic domains of parameter space.

\begin{figure}
\vskip 0.5in
\vspace*{-0.75in}
%\hspace*{-.70in}
\begin{minipage}{8in}
\epsfig{file=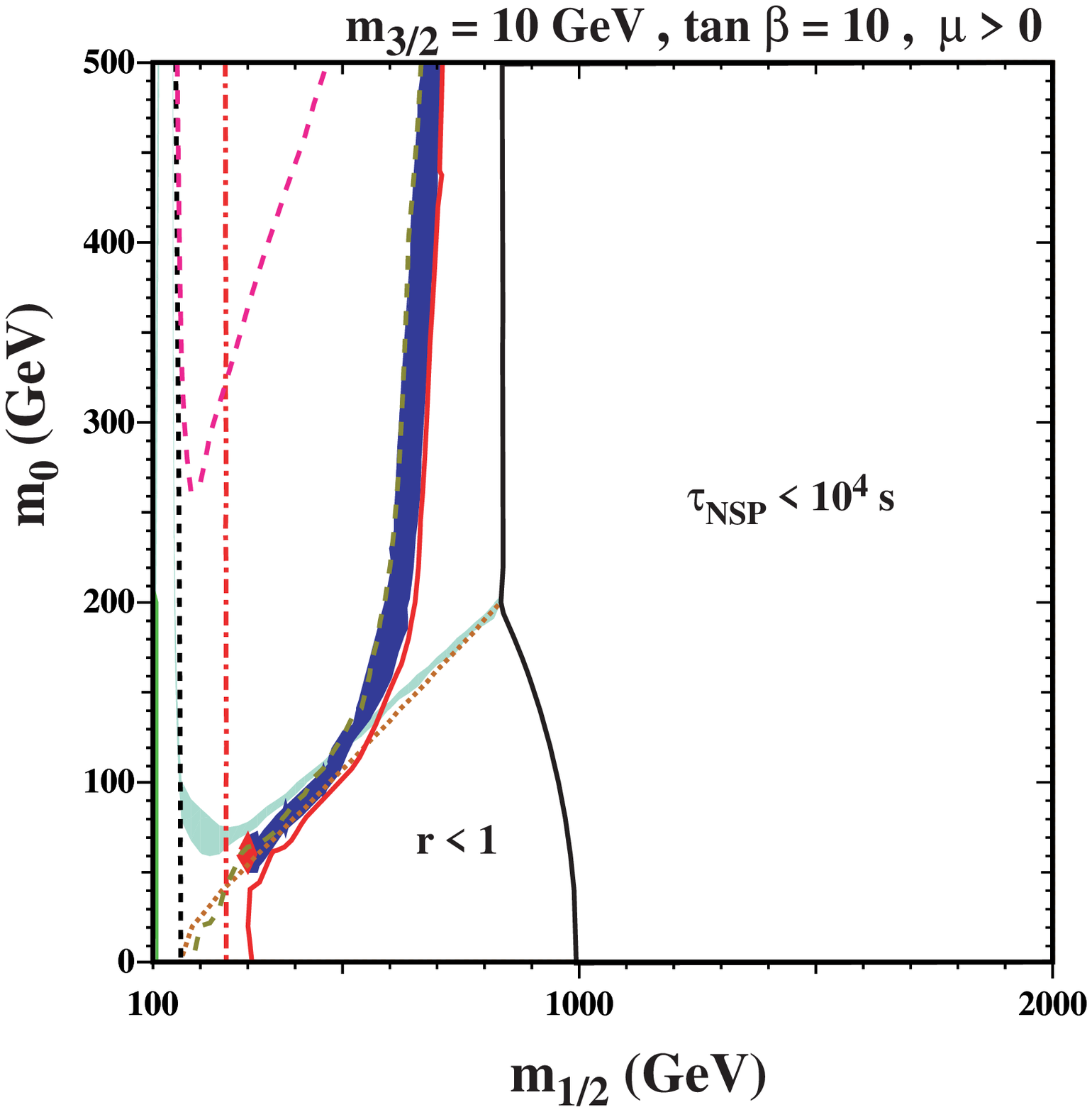,height=3.3in}
\hspace*{-0.17in}
\epsfig{file=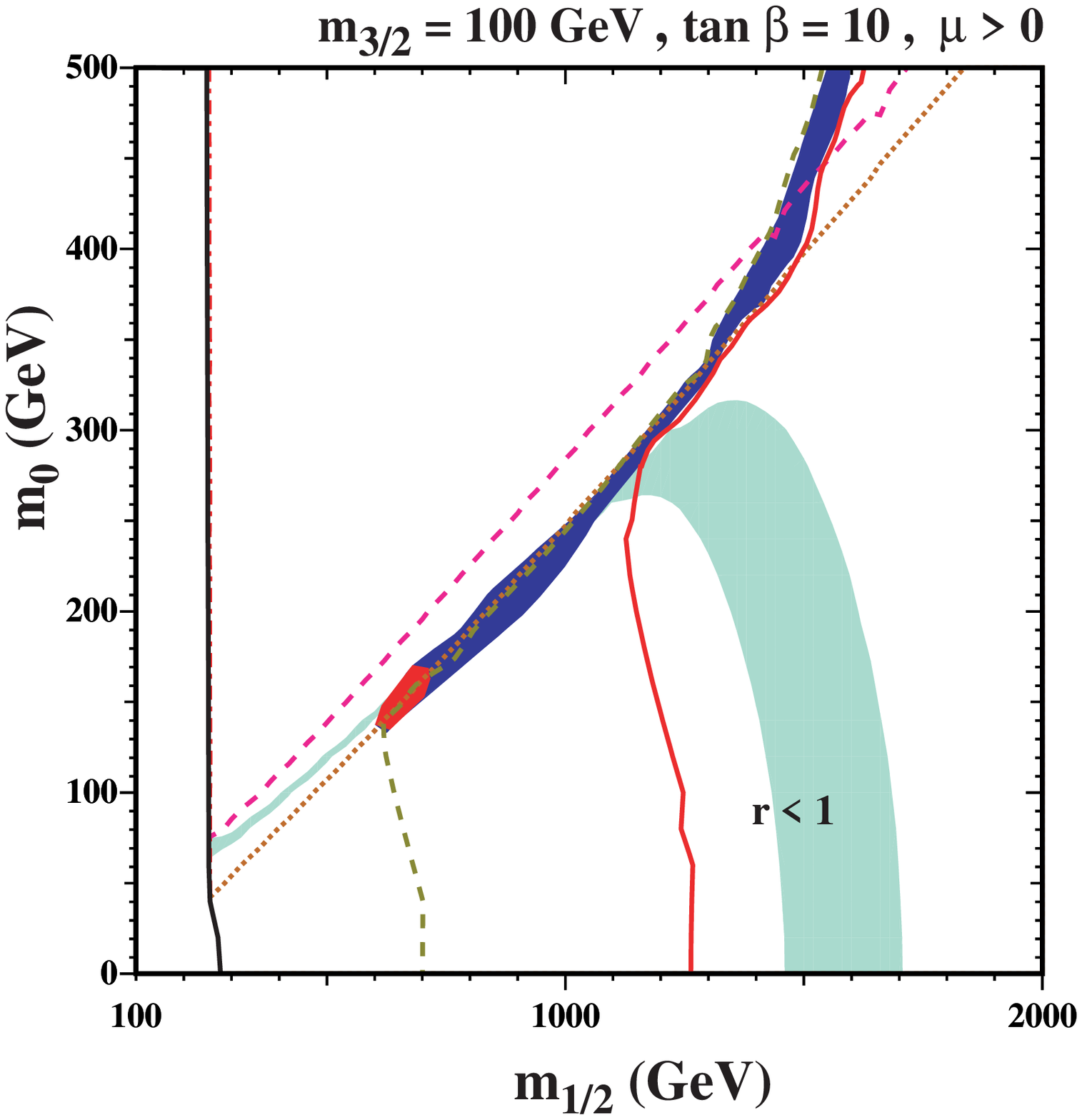,height=3.3in}
\hfill
\end{minipage}
\begin{minipage}{8in}
%\hskip -1.40in
%\vskip -.75in
\epsfig{file=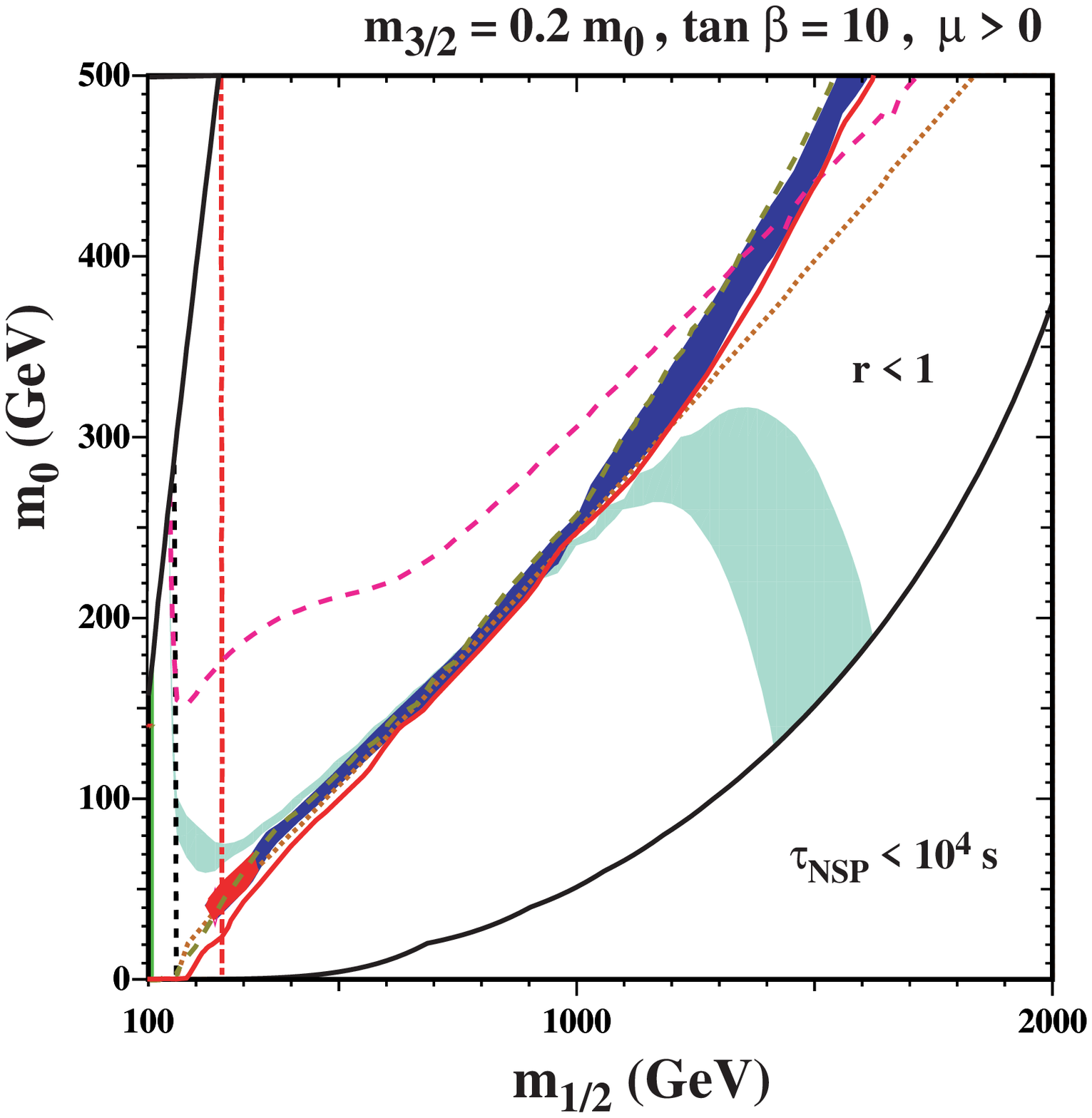,height=3.3in}
\hspace*{-0.2in}
\epsfig{file=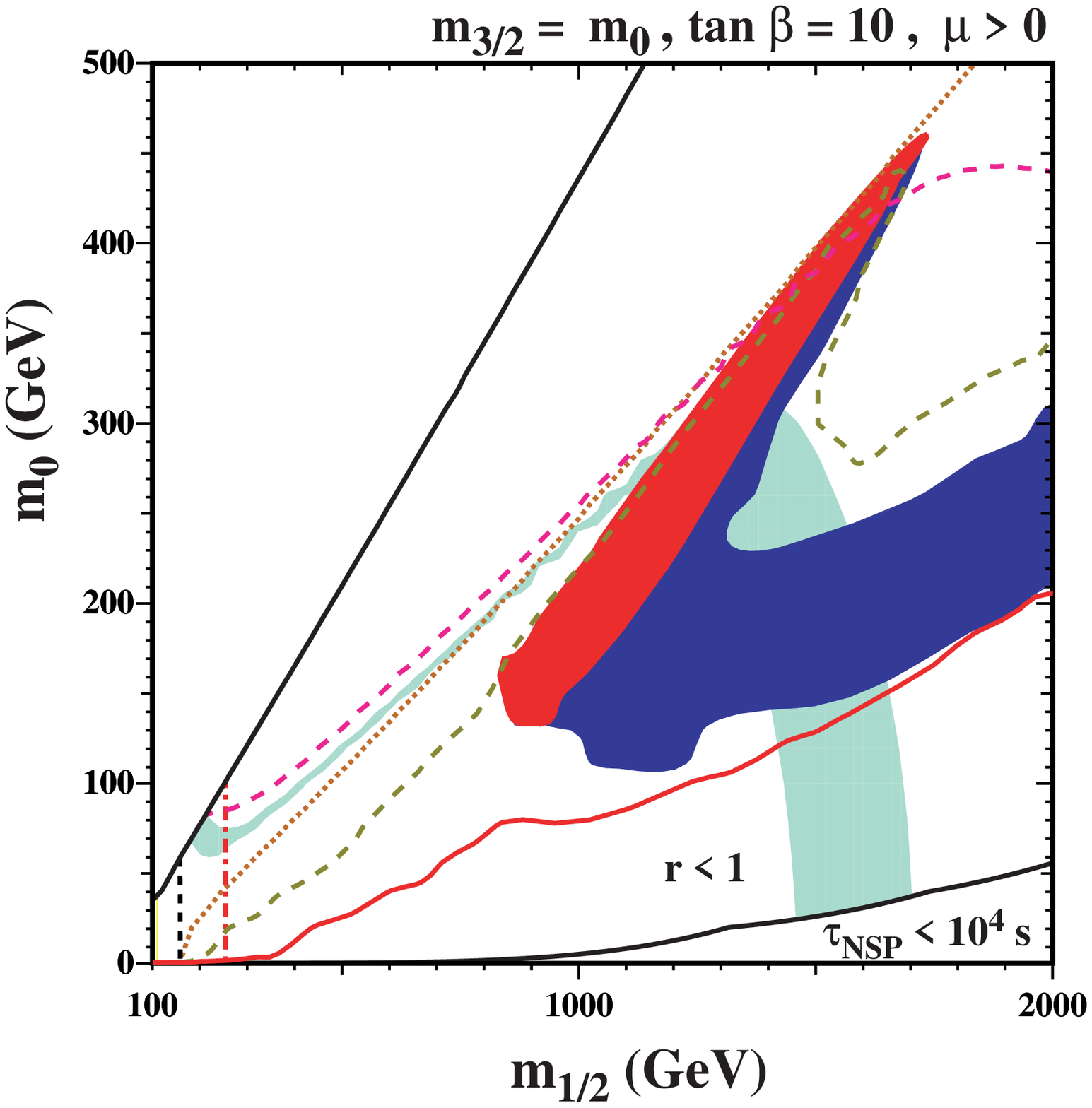,height=3.3in} \hfill
\end{minipage}
%\vskip 2.5in
\caption{
{\it 
The $(m_{1/2}, m_0)$ planes for $\mu > 0$, $\tan
\beta = 10$ and (a) $m_{3/2} = 10$~GeV, (b) $m_{3/2} = 100$~GeV, (c)  
$m_{3/2} = 0.2 m_0$ and (d) $m_{3/2} = m_0$. We restrict our attention 
to the regions between the solid black lines, where the gravitino is the 
LSP and the NSP lifetime exceeds $10^4$~s. In each panel, 
the near-vertical dashed black (dash-dotted red) line is the constraint
$m_{\chi^\pm} > 104$~GeV ($m_h > 114$~GeV), the upper (purple) dashed line 
is the
constraint $\Omega_{3/2} h^2 < 0.129$, and the light green shaded region
is that where the NSP would have had $0.094 \le \Omega h^2 \le
0.129$ if it had not decayed. The solid red (dashed grey-green) line is 
the region now (previously) allowed by the light-element abundances: $r < 
1$ as
described in the text. The red (blue) shaded region is that where the
\li7 abundance could have been improved by NSP decays, but which is now
excluded by the \he3 ({\rm D}) constraint.}}
\label{fig:CMSSM10} 
\end{figure}

In the first set of scenarios, shown in Fig.~\ref{fig:CMSSM10}, we fix the
ratio of supersymmetric Higgs vacuum expectation values $\tan \beta = 10$,
which is among the lower values consistent with our hypotheses, and assume
(a) $m_{3/2} = 10$~GeV, (b) $m_{3/2} = 100$~GeV, (c) $m_{3/2} = 0.2 m_0$
and (d) $m_{3/2} = m_0$. In each panel of Fig.~\ref{fig:CMSSM10}, we
display accelerator, astrophysical and cosmological constraints in the
corresponding $(m_{1/2}, m_0)$ planes, concentrating on the regions to the
right of the near-vertical black lines, where the gravitino is the LSP.
The vertical black dashed and (red) dot-dashed lines represent the 
lower limits on $m_{1/2}$ implied
for each value of $m_0$ by the non-observation of a chargino and a Higgs 
boson at LEP, the latter having a theoretical uncertainty $\delta m_{1/2} 
\sim 50$~GeV. The
(pale green) narrow diagonal strips represent the regions where the relic
density of the NSP would have lain in the range $0.094 \le \Omega h^2 \le
0.129$ favoured by WMAP and other measurements of the cold dark matter
density, {\it if the gravitino had} {\bf not} {\it been the LSP}. In fact,
the gravitino is always the LSP in the scenarios considered. The NSP may
be either the lightest neutralino $\chi$ or the lighter supersymmetric
partner of the $\tau$ lepton: this is lighter than the neutralino $\chi$
below the (red) dotted line. 

Below and to the right of the upper (purple) dashed lines, the density of
relic gravitinos produced in the decays of other supersymmetric particles
is always below the WMAP upper limit: $\Omega_{3/2} h^2 \le 0.129$. To the
right of the lower black solid lines, the lifetime of the next-to-lightest
supersymmetric particle (NSP) falls below $10^4$~s, and the analysis of
\cite{cefo} cannot evaluate the astrophysical constraints from the
light-element abundances, in the absence of a suitably modified BBN code.
The code used in \cite{cefo}, when combined with the observational
constraints used in~\cite{cefo}, yielded the astrophysical constraint
represented by the dashed grey-green lines in the different panels of
Fig.~\ref{fig:CMSSM10}.

These constraints on the CMSSM parameter plane were computed in
\cite{eoss5}. For each point in the $(m_{1/2},m_0)$, the relic density of
either $\chi$ or ${\tilde \tau}$ is computed and $\zeta_X$ is determined
using $\Omega_X h^2 = 3.9 \times 10^7$ GeV $\zeta_X$.  When $X = {\tilde
\tau}$, $\zeta_X$ is reduced by a factor of 0.3, as only 30\% of stau
decays result in electromagnetic showers which affect the element
abundances at these lifetimes. In addition, at each point, the lifetime of
the NSP is computed. Then for each $\tau_X$, the limit on $\zeta_X$ is
found from the results shown in Fig. \ref{fig:tauzetacleft}. The region to
the right of this curve where $r = \zeta_X/\zeta_X^{limit} < 1$ is
allowed.

The astrophysical constraints obtained with the newer abundance limits
used here yields the solid red lines in Fig.~\ref{fig:CMSSM10}. The
examples where $\tau_X$ and $\zeta_X$ for the NSP decays fall within the
ranges shown in Fig.~\ref{fig:ridiculous}(a), and hence are suitable for
modifying the $^7$Li abundance, are shown as red and blue shaded regions
in each panel of Fig.~\ref{fig:CMSSM10}. We see that they straddle the
erstwhile WMAP strips~\footnote{This raises the possibility that a
discovery of supersymmetry might have an ambiguous interpretation -
neutralino LSP or gravitino LSP - in the absence of supplementary
information.}. If we had been able to allow a Deuterium abundance as low
as D/H $\sim (1-2) \times 10^{-5}$, the blue shaded region would have been
able to resolve the Li discrepancy in the context of the CMSSM with
gravitino dark matter. The blue region that we now regard as excluded by
the lower limit on D/H, which is stronger than that used in~\cite{cefo},
extends to large $m_{1/2}$. The red shaded region, which is consistent
even with this limit on D/H, but yields very large $^3$He/D, is close to
the Higgs lower limit on $m_{1/2}$ for small $m_{3/2}$, moving to larger
$m_{1/2}$ for larger $m_{3/2}$, so as to keep $\tau_{NSP}$ within the
desired range.

We displayed in Fig.~\ref{fig:tauzetacleft} the impact of the improved
lower limit on D/H and the new $^3$He constraint on the abundance of an
unstable particle, as a function of its lifetime. Interpreting this as a
constraint on NSP decay into a gravitino, the panels in
Fig.~\ref{fig:CMSSM10} show as solid red lines the additional
restrictions these constraints impose on the $(m_{1/2}, m_0)$ planes for
different values of $m_{3/2}$. The effects for small $m_{3/2} = 10$~GeV
[in panel (a)] $m_{3/2} = 100$~GeV when $m_0$ is large [in panel (b)] and
$0.2 m_0$ [in panel (c)] are relatively modest.  This is because the limit
occurs in a region where the NSP is a neutralino, and the relic density
varies relatively rapidly. Hence a small change in the $m_{1/2}$ or $m_0$
results in a large change in $\zeta_X$, and the old and new bounds are
relatively close. However, they do bite in the neighbourhood of the shaded
$^7$Li blobs, and have the effect of excluding them entirely. The effects
for large $m_{3/2} = 100$~GeV and small $m_0$ [in panel (b)] and $m_{3/2}
= m_0$ [in panel (d)] are relatively large, mainly due to the slow
variation of $\zeta_X$ near the limit which is characteristic of the
${\tilde \tau}$ NSP region, as can be seen from the relatively wide WMAP
strips in this region. Reflecting this wide separation between the (old)
dashed grey-green lines and the (new) solid red lines, we see large
red and blue swaths in panel (d), where the $^7$Li abundance could have
been reduced, but the $^3$He/D and/or D/H ratios are
unacceptable~\footnote{We note in passing that in several panels there are
parts of the blue and/or red regions at large $m_0$ that are also excluded
by the WMAP relic density limit.}.

\begin{figure}
\vskip 0.5in
\vspace*{-0.75in}
%\hspace*{-.70in}
\begin{minipage}{8in}
\epsfig{file=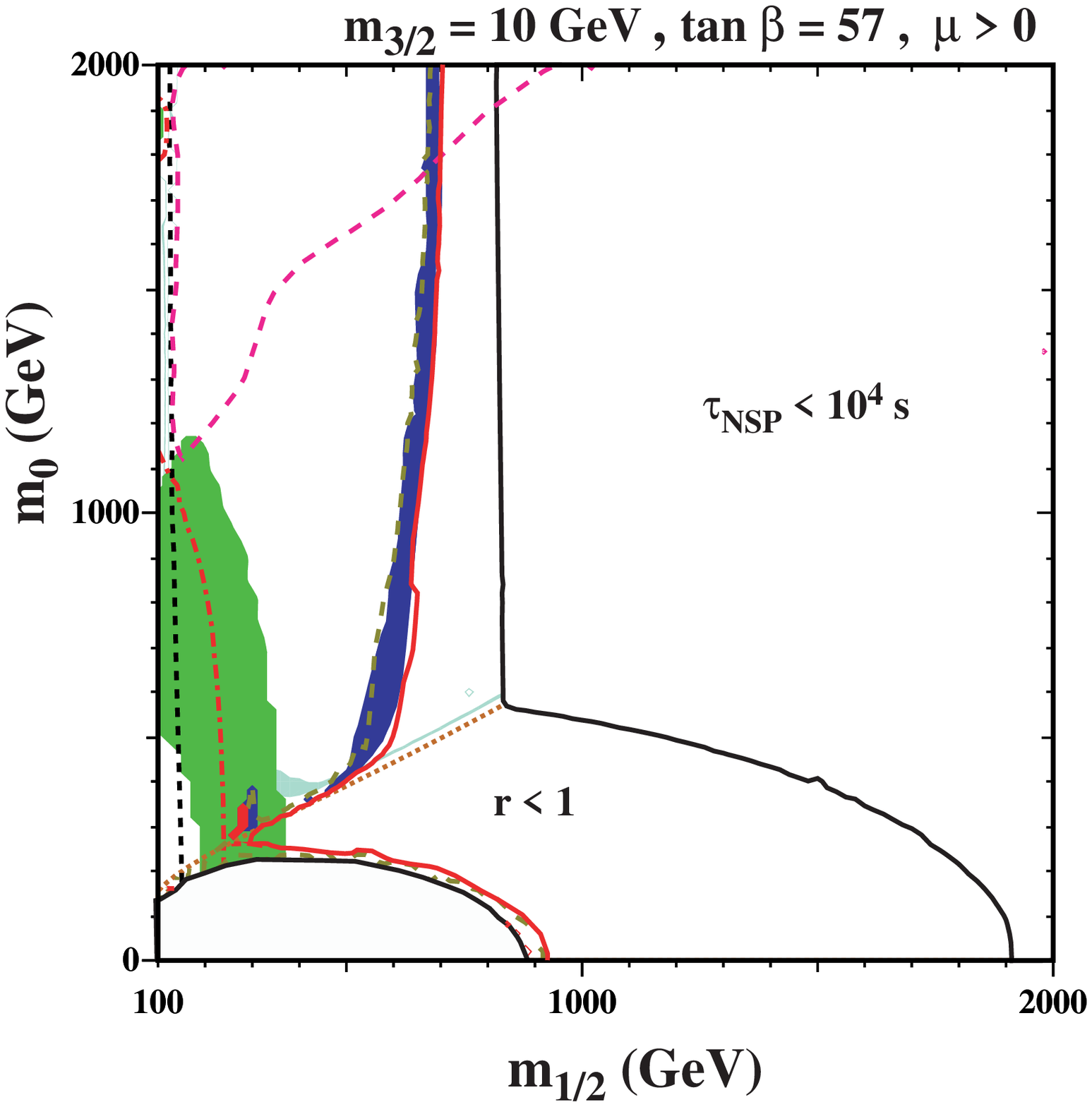,height=3.3in}
\hspace*{-0.17in}
\epsfig{file=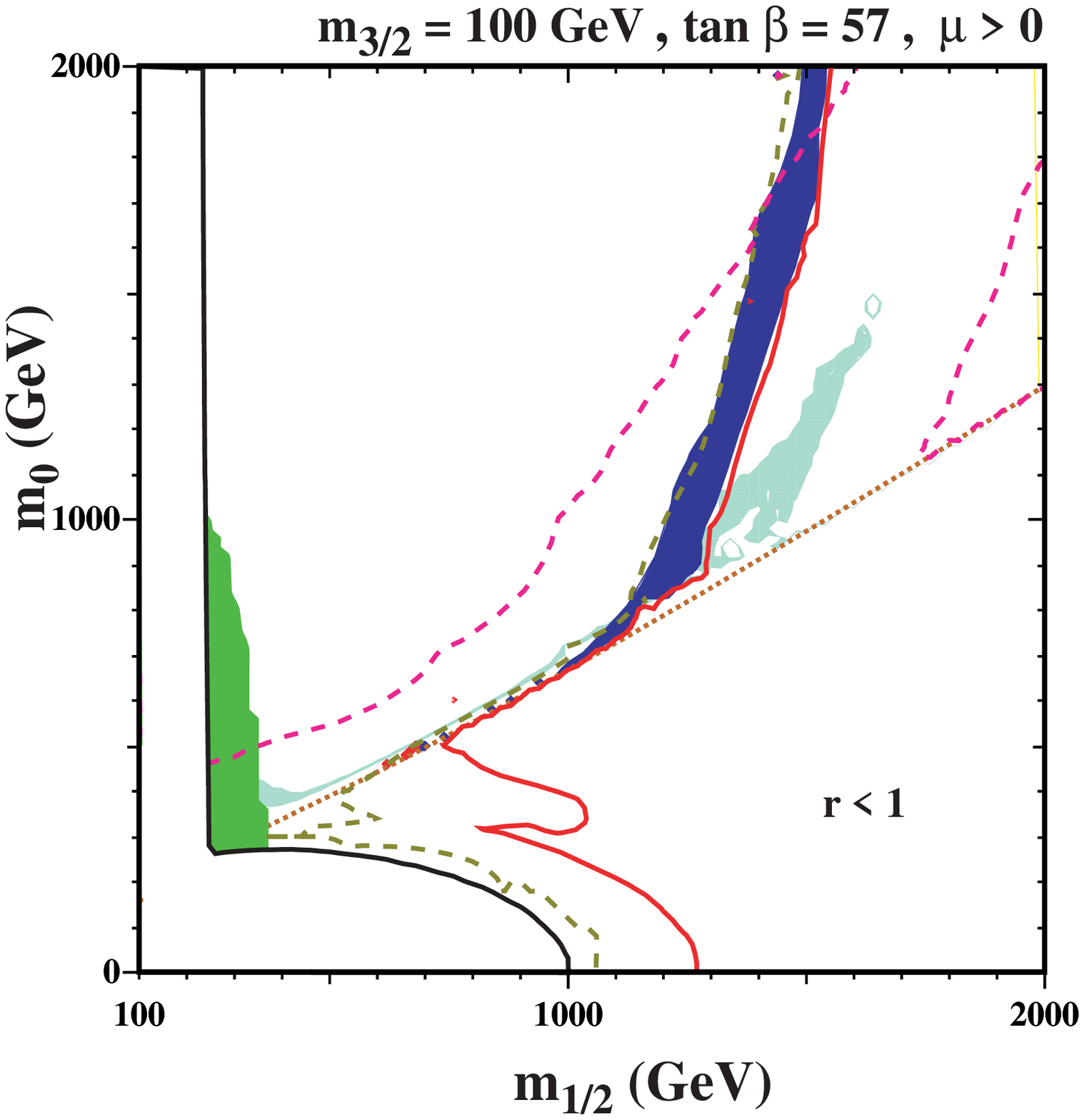,height=3.3in}
\hfill
\end{minipage}
\begin{minipage}{8in}
%\hskip -1.40in
%\vskip -.75in
\epsfig{file=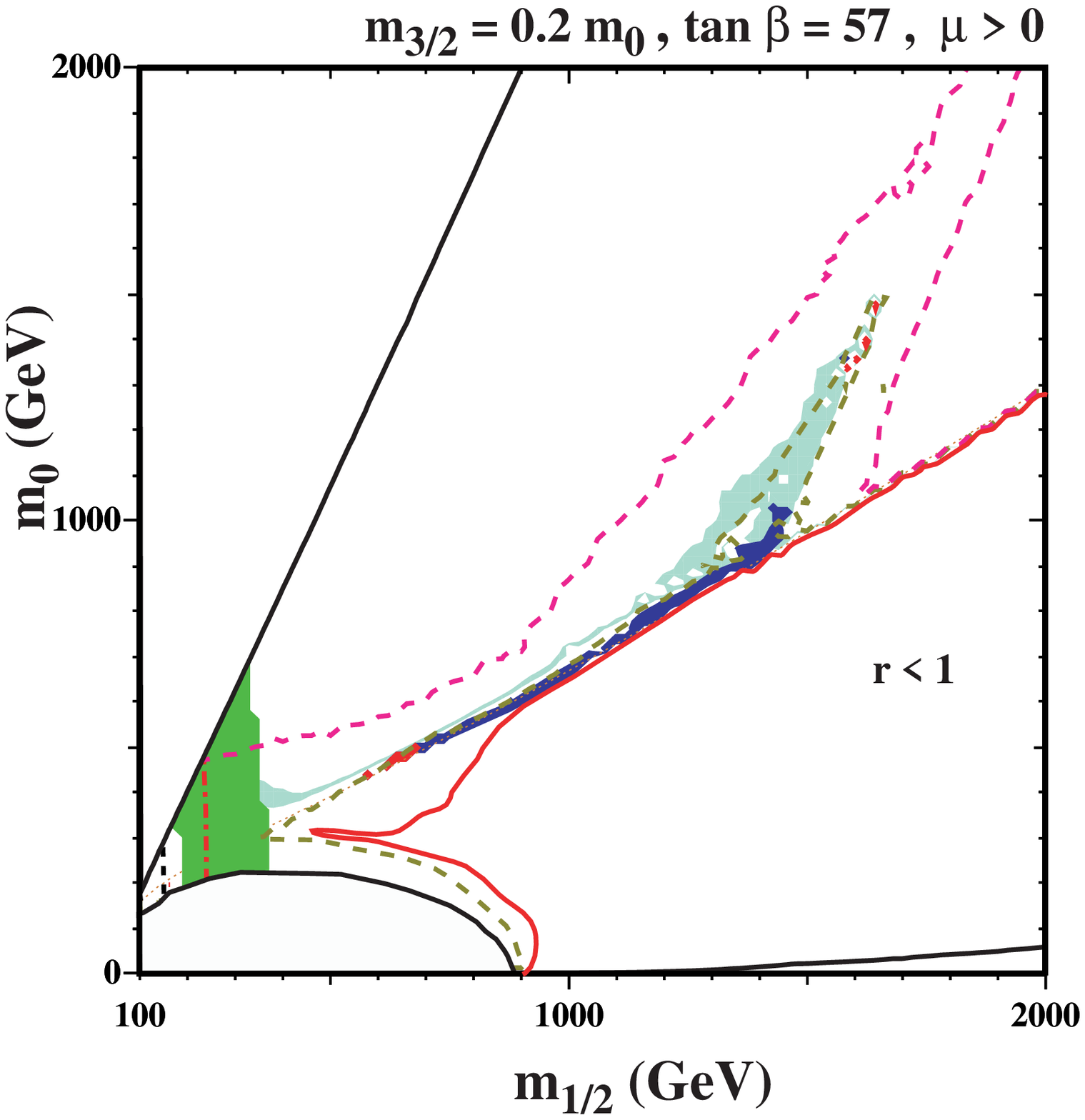,height=3.3in}
\hspace*{-0.2in}
\epsfig{file=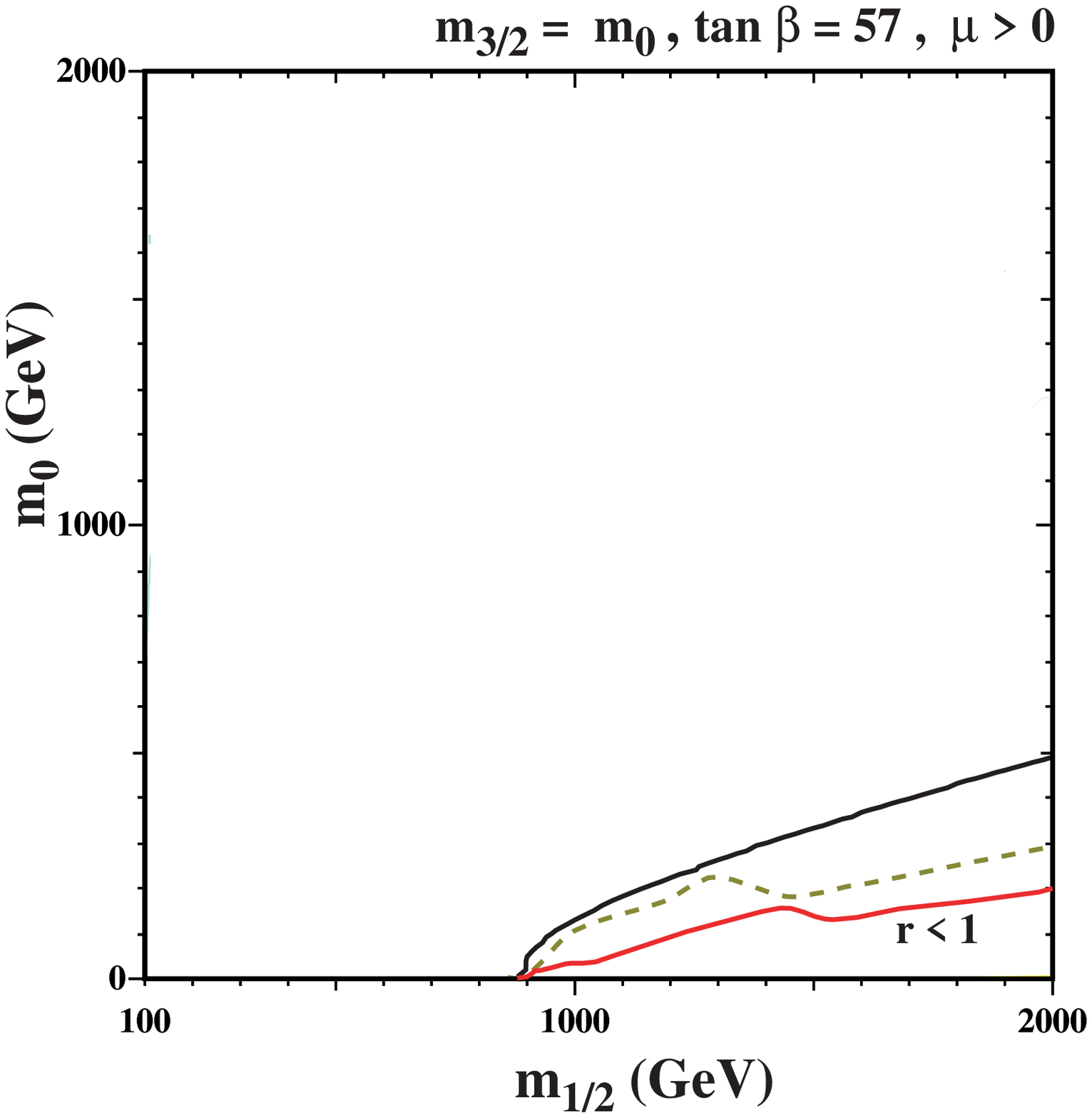,height=3.3in} \hfill
\end{minipage}
%\vskip 2.5in
\caption{
{\it
As in Fig.~\ref{fig:CMSSM10}, but now for $\tan \beta = 57$.}}
\label{fig:CMSSM50} 
\end{figure}

Fig.~\ref{fig:CMSSM50} shows the corresponding $(m_{1/2}, m_0)$ planes for
the choice $\tan \beta = 57$, which is among the larger values allowed in the context
of the constrained MSSM. Although the
shapes of the allowed regions are rather different from the previous $\tan
\beta = 10$ case, the qualitative conclusions are similar. The $^3$He
constraint again has relatively modest impact. However, in all cases the
red shaded $^7$Li regions are excluded by the $^3$He constraint, and the
blue regions by the D abundance. We note that, in models with $m_{3/2} = 
m_0$ as in Fig.~\ref{fig:CMSSM50}(d),
very little of the $(m_{1/2}, m_0)$ plane admits gravitino dark matter 
and, in the viable corner, there were no possibilities for depleting \li7.  

\begin{figure}
\vskip 0.5in  
\vspace*{-0.75in}
%\hspace*{-.70in}
\begin{minipage}{8in}
\epsfig{file=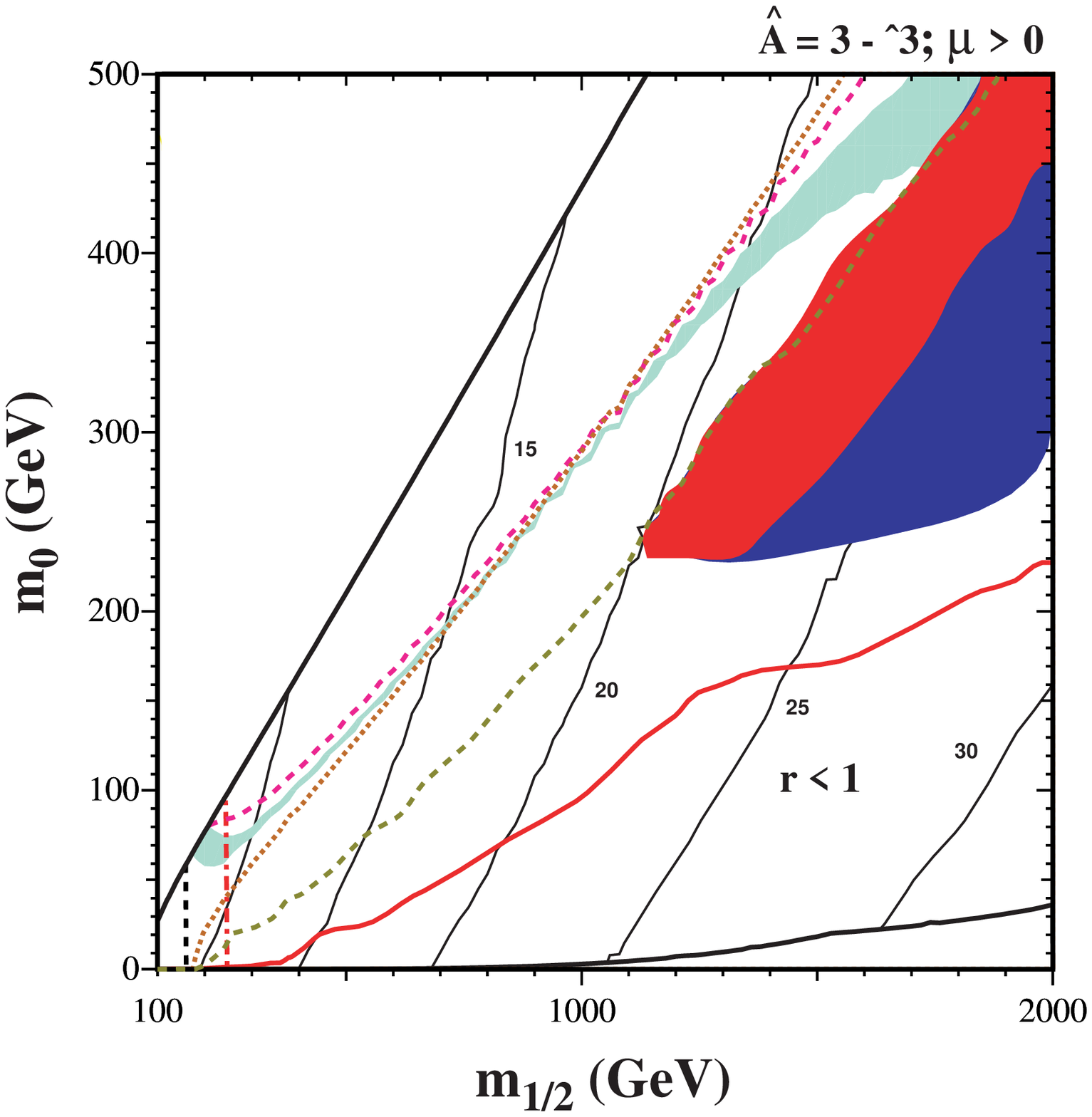,height=3.3in}
\hspace*{-0.17in}
\epsfig{file=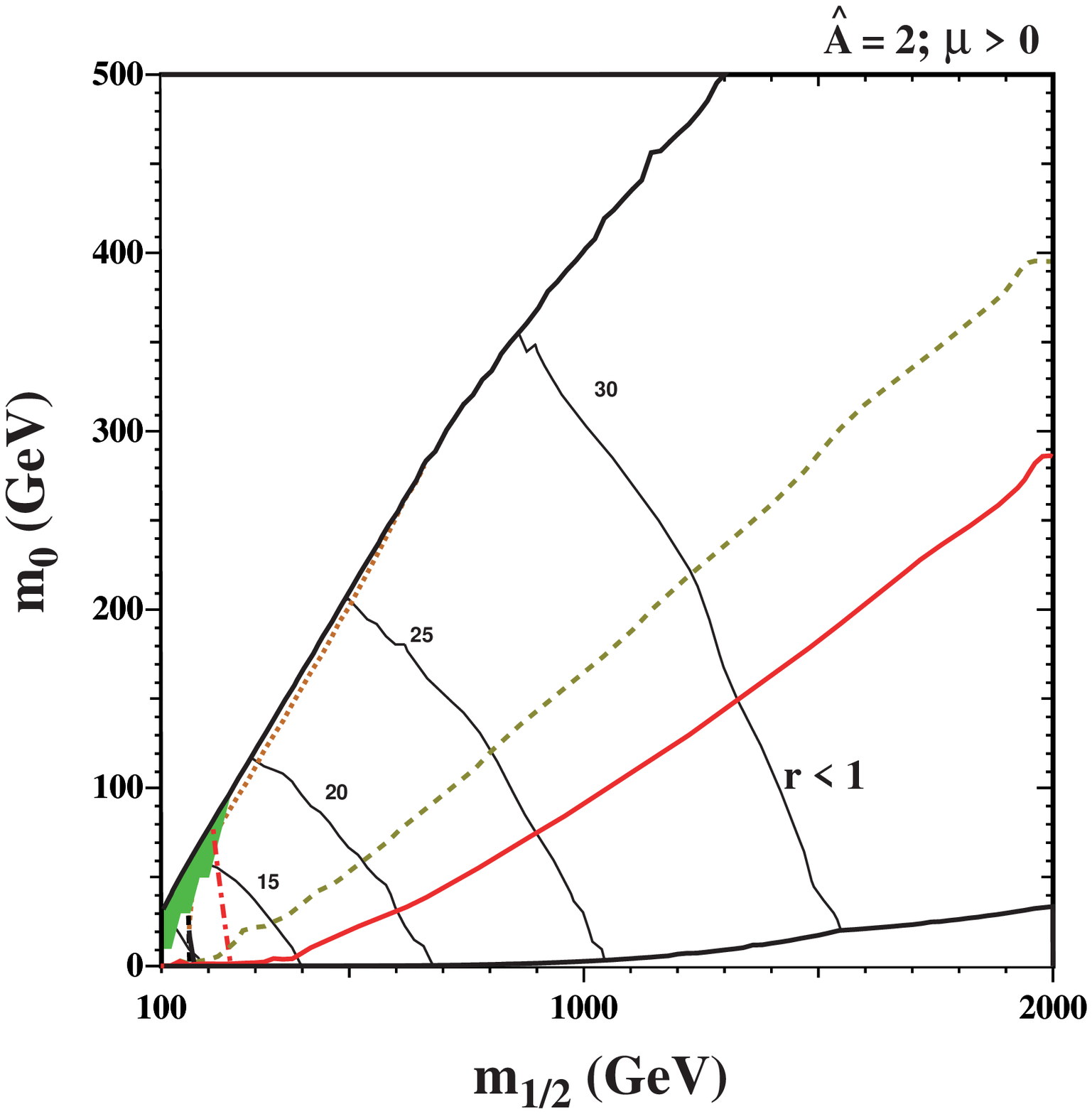,height=3.3in}
\hfill
\end{minipage}
\caption{
{\it
As in Fig.~\ref{fig:CMSSM10}, but now for very constrained models 
motivated by mSUGRA. The value of $\tan \beta$ is fixed by the vacuum 
conditions, and varies across the $(m_{1/2}, m_0)$ planes with values 
indicated by the steep black contours. These models are 
specified by the choices of $A \equiv {\hat A} m_{3/2}$: (a) ${\hat A} = 
(3 - \sqrt{3})$, the Polonyi model, and (b) ${\hat A} = 2$.}} 
\label{fig:mSUGRA}
\end{figure}

Finally, we consider very constrained models motivated by minimal
supergravity (mSUGRA), in which not only is $m_{3/2} = m_0$, but also the
trilinear soft supersymmetry-breaking parameter $A$ determines the
bilinear soft supersymmetry-breaking Higgs-mixing parameter: $B = A -
m_{3/2}$. This is compatible with the values of $\mu$ and $m_A$ specified
by the electroweak vacuum conditions for only one value of $\tan \beta$
for any given pair of values of $(m_{1/2}, m_0)$ \cite{vcmssm}. Such
mSUGRA models are then specified by a choice of $A \equiv {\hat A}
m_{3/2}$: the panels of Fig.~\ref{fig:mSUGRA} assume (a) ${\hat A} = (3 -
\sqrt{3})$ as found in the simple Polonyi model of supersymmetry breaking
in mSUGRA \cite{pol}, and (b) ${\hat A} = 2$.  For ${\hat A} = (3 -
\sqrt{3})$, the contours of constant $\tan \beta$ are approximately
vertical, and range from about 10 at low $m_{1/2}$ to about 30 at high
values of $m_{1/2}$.  In the interesting region of panel (a) where \li7
can be depleted, $\tan \beta \sim 20 - 30$.
As before, we consider here only regions of the $(m_{1/2},
m_0)$ planes between the two solid black lines: above the higher one, the
gravitino is no longer the LSP, and below the lower one the lifetime falls
below $10^4$~s. In addition to the constraints discussed earlier, panel
(b) also displays a small green shaded region at low $m_{1/2}$ that is
excluded by $b \to s \gamma$ decay.

We see that there is a large difference between the effects of
implementing the old and new light-element constraints in panel (a) of
Fig.~\ref{fig:mSUGRA} for ${\hat A} = (3 - \sqrt{3})$, the Polonyi value,
whereas the effect in panel (b) for ${\hat A} = 2$ is smaller. In the
Polonyi case, there are large \li7-friendly regions that are excluded by
the \he3 and D constraints. This reflects the fact that $\tan \beta$ is
relatively small in this case, so the model is qualitatively similar to 
the $m_{3/2} = m_0$ case for $\tan \beta = 10$ shown in panel (d) of
Fig.~\ref{fig:CMSSM10}. On the other hand, $\tan \beta$ is typically
larger for ${\hat A} = 2$, and when combined with a 
smaller stau mass, we find  no visible \li7-friendly region,
as a result of small yet significant shifts in the values of both 
$\zeta_X$ and
$\tau_X$.

\section{Conclusions}

In the absence of a convincing astrophysical explanation for the apparent
discrepancy between the observed abundance of \li7 and that calculated on
the basis of the baryon-to-photon ratio inferred from CMB observations,
it has been natural to explore the possible effects of late-decaying
massive particles. Indeed, they could suppress the primordial abundance of
\li7, but at some price. Either the abundance of Deuterium should be very
low compared with the latest available measurements, and/or the primordial
\he3/D ratio must have been very high.

The latest observations of remote cosmological clouds along the lines of 
sight of high-redshift quasars suggest that D/H $> 2.2 \times 10^{-5}$, 
ruling out much of the parameter space for unstable particles that would 
otherwise have been suitable for diminishing the \li7 abundance to agree 
with observations. A significant part of this parameter space would have 
been allowed by the more relaxed limit D/H $> 1.3 \times 10^{-5}$ 
considered previously~\cite{cefo}.

The remaining part of the parameter space for unstable particles that is 
consistent with the current lower limit would yield a \he3/D ratio at 
least an order of magnitude higher than the proto-solar value. Since D has 
been destroyed by stars, reducing its abundance from approximately
2.5 to 1.5 $\times 10^{-5}$, while the \he3 abundance is thought to have 
remained roughly constant, a primordial ratio of \he3/D $> 1$ or 2 is 
unacceptable. This closes the remaining loophole for suppressing \li7 
without running into conflict with the other light-element abundances.

We have also analyzed the potential bounds imposed by the \li6 abundance.
In principle,  the high growth of its abundance between BBN and the
formation of halo stars could also have been explained by the decays of the NSP
in supersymmetric models, were it not for either the low resulting 
abundance of D/H or the high ratio of \he3/D.

So what is the interpretation of the apparent discrepancy between the
calculations of the primordial \li7 abundance and the Spite plateau? We
have argued that the origin of the discrepancy cannot be the possible
existence of unstable particles able to destroy the primordial nucleus.
This leaves the problem open. Systematic uncertainties in nuclear effects
such as higher $^7$Be + D reaction rates have been considered~\cite{coc,cfo4},
but seem unable to modify substantially the abundance of \li7. 
Stellar mechanisms of depletion may be the last resort \cite{dep}. Perhaps
other new and exciting astrophysical or physical effects will  have to
be considered.  

A significant output of this analysis has been the demonstration of the
importance of the \he3/D constraint on late-decaying massive
particles as argued in \cite{kawa}. Can one make the \he3 constraint more precise? This would
require considering in more detail the cosmic evolution of D and \he3. One
should allow for the possibility of exotic effects such as large-scale
destruction of this isotope in primitive structures such as massive
Population-III stars, followed by moderate production by normal galactic
evolution. However, any such scenario should consider simultaneously the
cosmic evolution of D. As it is more fragile than \he3, D would also be
destroyed in any Population-III stars. As we have seen, the \he3
constraint sharpens the embarrassment of supersymmetric models with heavy
particles whose lifetimes are $\sim 10^7$ to $10^8$~s. This interest
motivates more detailed studies of the cosmic evolution of \he3 and D, as
well as \li6 and \li7.

Finally, it would be very useful if the observed abundances of D/H in
quasar absorption systems were improved.  There is currently considerable
dispersion in the observed abundances.  We note that two such systems in
the directions of Q2206-199 at $z=2.0762$ with D/H = 1.65 $\pm 0.35 \times
10^{-5}$~\cite{pb} and PKS1937-1009 at $z=3.256$, with D/H =
1.6$^{+0.25}_{-0.30} \times 10^{-5}$~\cite{cwof} have quite low abundances
of D, similar to that observed at present in the ISM of our
Galaxy~\cite{moos}. If these measurements were to represent the correct
D/H abundance in those clouds, \li7 depletion and \li6 production by
sparticle decay would be a viable option, though by solving one problem we
would open two new problems. Why is the D/H abundance in most of the other
absorption systems significantly higher, and how can we account for the
D/H abundances in the solar system, which are also in the range 1.5 - 2.5
$\times 10^{-5}$?

\section*{Acknowledgments}

We thank R. Cyburt, B. Fields, Y. Santoso and V. Spanos
for collaborations on related topics.
The work of K.A.O. and E.V. was supported by the Project "CNRS/USA", and
the work of K.A.O. was also supported partly by DOE grant
DE--FG02--94ER--40823.

\end{document}